\documentclass[aps,prl, reprint,superscriptaddress,longbibliography,nobibnotes]{revtex4-2}
\usepackage{amsfonts,amscd,mathrsfs,amsmath,amsthm,amssymb}
\usepackage{mathtools}
\usepackage{xparse}
\usepackage{graphicx}
\usepackage{float}
\usepackage{pifont}
\usepackage[dvipsnames]{xcolor}
\usepackage{colortbl}
\usepackage{multirow}
\usepackage{enumerate}
\usepackage{comment}
\usepackage{booktabs}
\usepackage{cancel}
\usepackage{subcaption}
\IfFileExists{old-arrows.sty}{\usepackage[new]{old-arrows}}{}
\usepackage{circuitikz}
\IfFileExists{lieart.sty}{\usepackage{lieart}}{}
\usepackage{listings}
\usepackage{xcolor}
\definecolor{codegreen}{rgb}{0,0.6,0}
\definecolor{codegray}{rgb}{0.2,0.2,0.2}
\definecolor{codepurple}{rgb}{0.58,0,0.82}
\definecolor{backcolour}{rgb}{0.95,0.95,0.92}
\lstdefinestyle{mystyle}{
 backgroundcolor=\color{backcolour},
 commentstyle=\color{codegreen},
 keywordstyle=[1]\color{magenta},
 keywordstyle=[2]\color{blue},
 otherkeywords = {PerfectGroup,CharacterTable,Irr,Norm,ScalarProduct,gap>},
 morekeywords = [1]{PerfectGroup,CharacterTable,Irr,Norm,ScalarProduct,Degree},
 morekeywords = [2]{gap>,>},
 numberstyle=\tiny\color{codegray},
 stringstyle=\color{codepurple},
 basicstyle=\ttfamily\footnotesize,
 breakatwhitespace=false,
 breaklines=true,
 captionpos=b,
 keepspaces=true,
 numbers=left,
 numbersep=5pt,
 frame = single,
 showspaces=false,
 showstringspaces=false,
 showtabs=false,
 morecomment=[l][\color{codegreen}]{\#},
 alsoletter=?>,
 tabsize=2
}
\IfFileExists{physics.sty}{\usepackage{physics}}{%
 \providecommand{\norm}[1]{\left\lVert##1\right\rVert}%
}
\usepackage{bm}
\IfFileExists{bbm.sty}{\usepackage{bbm}}{%
}
\usepackage{caption}
\PassOptionsToPackage{hyphens}{url}\usepackage{hyperref}
\hypersetup{
 colorlinks=true,
 linkcolor=magenta!80!black,
 citecolor=magenta!80!black,
 breaklinks=true,
}
\usepackage[capitalise]{cleveref}

\theoremstyle{plain}

\theoremstyle{definition}

\newtheorem*{conjecture*}{Conjecture}

\AddToHook{cmd/appendix/before}{%
 \setcounter{equation}{0}%
 \setcounter{theorem}{0}%
 \setcounter{lemma}{0}%
}

\providecommand{\SU}{}

\providecommand{\C}{}

\DeclareMathAlphabet{\mathsfit}{T1}{\sfdefault}{\mddefault}{\sldefault}
\SetMathAlphabet{\mathsfit}{bold}{T1}{\sfdefault}{\bfdefault}{\sldefault}

\renewcommand{\SU}{\mathrm{SU}}

\renewcommand{\C}{\mathcal{C}}

\newcommand{\wt}{\operatorname{wt}}
\renewcommand{\Tr}{\operatorname{Tr}}
\newcommand{\calC}{\mathcal C}

\begin{document}

\title{Time-Reversal Selection Rules for Quantum Error Correction}
\thanks{These authors contributed equally to this work.}

\author{Eric Kubischta}
\email{emk25g@fsu.edu}
\affiliation{Quantum Initiative, Florida State University, Tallahassee, FL 32306}
\affiliation{Department of Mathematics, Florida State University, Tallahassee, FL 32306}

\author{Ian Teixeira}
\email{iteixeira@ucsd.edu}
\affiliation{Department of Mathematics, University of California, San Diego, CA 92093}

\begin{abstract}
We apply time-reversal symmetry to quantum codes and show that it imposes parity selection rules on the physical error algebra. A time-reversal-invariant logical qubit on an odd number of spins is a Kramers doublet, forcing every even-weight Pauli to act as a scalar. Consequently, all even-weight Knill--Laflamme conditions hold automatically, so single-qubit error detection implies correction. We then reinterpret the Rains shadow enumerator through time reversal: each coefficient is a sum of error-resolved overlaps between a code and its time-reversed image. 
\end{abstract}

\maketitle

\emph{Introduction.---}
Time reversal entered quantum mechanics as the deceptively simple operation of running a physical process backward: momenta and angular momenta reverse, while positions do not. In quantum theory, however, time reversal is fundamentally unlike an ordinary unitary symmetry. Wigner showed that it is represented by an antiunitary operator, and that for a system with half-integer total spin this operator squares to $-1$ \cite{Wigner}. From this algebraic fact Kramers obtained one of the earliest and most enduring examples of symmetry protection in quantum mechanics: every energy level of a time-reversal-invariant system with half-integer spin is at least doubly degenerate \cite{Kramers}. The theorem requires no continuous symmetry, no fine tuning, and no detailed knowledge of the Hamiltonian beyond its invariance under time reversal.

Kramers degeneracy is only the spectral manifestation of a broader selection rule. The two states of a Kramers pair are exchanged by time reversal, and time-reversal-even observables cannot distinguish or mix them arbitrarily. This restriction has become a recurring organizing principle throughout quantum physics. It governs degeneracies and transition amplitudes in atomic and molecular spectroscopy, determines the symplectic universality class of spin-orbit-coupled systems \cite{Dyson}, and protects the helical boundary states of time-reversal-invariant topological phases \cite{KaneMele,FuKaneMele}. In each case, the underlying mechanism is the same: antiunitarity divides the operator algebra into time-reversal-even and time-reversal-odd sectors, and the Kramers structure sharply constrains how those sectors can act within a protected doublet.

Quantum error correction appears to demand a fundamentally different kind of protection. A quantum code is not required merely to remain degenerate under a particular symmetry-preserving Hamiltonian. The Knill--Laflamme conditions require every product $E_a^\dagger E_b$ drawn from a prescribed error set to act as a scalar on the codespace \cite{KnillLaflamme}. This is an algebraic condition on an entire family of operators, rather than a spectral statement about a single Hamiltonian. For this reason, Kramers degeneracy and quantum error correction have generally been treated as distinct mechanisms: the former protects an energy doublet against symmetry-preserving perturbations, whereas the latter protects encoded information against a specified error set.

Despite the central role of time reversal elsewhere in quantum physics, its selection rule has not been applied to quantum error correction. Here we show that, for systems of spin-$1/2$ particles, the apparent gap between these two mechanisms disappears. Physical time reversal reverses every spin and therefore assigns a definite time-reversal parity to every Pauli string: a Pauli string is time-reversal even or odd precisely when its weight is even or odd. If a two-dimensional codespace on an odd number of qubits is invariant under time reversal, then the encoded qubit is itself a Kramers doublet. The ordinary Kramers selection rule consequently acts not merely on a Hamiltonian, but on the full physical error algebra.

This observation produces an exact operator-level selection rule. Every even-weight Pauli operator acts as a scalar on a time-reversal-invariant logical qubit, while every odd-weight Pauli operator acts as a traceless logical operator. Time reversal therefore supplies, identically and without optimization, every even-weight Knill--Laflamme condition. What ordinarily appears as spectral protection of a degenerate pair becomes exact error protection once the time-reversal grading is identified with Pauli-weight parity.

The immediate consequence is striking. For a time-reversal-invariant logical qubit, detection of all single-qubit Pauli errors already implies correction of an arbitrary single-qubit error. Error correction normally requires control not only of the individual errors $E_a$, but also of all cross terms $E_a^\dagger E_b$. When $E_a$ and $E_b$ are supported on different qubits, these cross terms have weight two. They are therefore time-reversal even and automatically act as scalars on the code. More generally, in constructing a distance-$2t+1$ code, one need impose only the odd-weight conditions through weight $2t-1$; every even-weight condition through weight $2t$ follows from symmetry. In particular, every time-reversal-invariant logical qubit has odd distance. These conclusions require neither a stabilizer description nor additivity, transversality, or a protecting parent Hamiltonian. They are intrinsic properties of the encoded subspace.

The same perspective reveals that time reversal was already present, in disguised form, in one of the fundamental invariants of quantum coding theory. Rains defined the quantum shadow enumerator by applying a sitewise ``spin flip'' to the code projector \cite{RainsShadow}. For qubits, we show that the spin-flipped projector is exactly the projector onto the time-reversed code. Every shadow coefficient is therefore an error-resolved overlap between a code and its time-reversed image. Rains-even codes are precisely time-reversal-invariant codes, Rains-odd codes are orthogonal to their time reverses.

\emph{Kramers selection rule.---}
Let $K$ be complex conjugation in the computational basis. The canonical time reversal of $n$ spin-$1/2$ degrees of freedom is
\begin{equation}
 \Theta=(iY K)^{\otimes n},\qquad \Theta^2=(-1)^n.
 \label{eq:theta}
\end{equation}
For abstract qubits, the same antiunitary is the pseudoreal, or charge-conjugation, structure of the defining $\SU(2)$ representation. Since each nonidentity Pauli changes sign,
\begin{equation}
 \Theta E\Theta^{-1}=(-1)^{\wt(E)}E
 \label{eq:pauliparity}
\end{equation}
for every Hermitian Pauli string $E$.

Let $P$ project onto a two-dimensional code $\calC$ and suppose $\Theta P\Theta^{-1}=P$, meaning the code $\calC$ is invariant under time-reversal symmetry. When physical errors $E$ are compressed to the codespace, they have the form
\begin{equation}
 PEP\big|_{\calC}=a_0 I_L+\bm a\cdot\bm\sigma_L.
 \label{eq:logicaldecomp}
\end{equation}
Logical time reversal fixes $I_L$ and reverses all three logical Pauli operators. Furthermore, we can decompose any Hermitian operator into time-reversal parity components,
\begin{equation}
 O_\pm=\frac12\left(O\pm\Theta O\Theta^{-1}\right).
 \label{eq:opparity}
\end{equation}
Invariance of $P$ lets time reversal pass through the compression. Comparing the coefficients of $PO_\pm P$ before and after logical time reversal gives
\begin{equation}
 PO_+P=\frac{\Tr(PO_+)}2P,
 \qquad \Tr(PO_-P)=0.
 \label{eq:generalrule}
\end{equation}
Equation~\eqref{eq:pauliparity} now turns this general Kramers rule into the Pauli-weight selection rule
\begin{equation}
 \boxed{
 \begin{aligned}
  \wt(E)\ \mathrm{even}&:\quad PEP=c_E P,\quad c_E=\tfrac12\Tr(PE),\\
  \wt(E)\ \mathrm{odd}&:\quad PEP=\bm a\cdot\bm\sigma_L,\quad \Tr(PEP)=0.
 \end{aligned}}
 \label{eq:selection}
\end{equation}
This rule has an exact rank-one companion. When $n$ is even, $\Theta^2=+1$ permits an invariant state $P=|\psi\rangle\!\langle\psi|$ with $\Theta|\psi\rangle=|\psi\rangle$; a Hermitian time-reversal-odd operator then equals its own negative in expectation, so
\begin{equation}
 \langle\psi|E|\psi\rangle=0
 \quad [\wt(E)\ \mathrm{odd}].
 \label{eq:rankone}
\end{equation}
This stores no logical information, but arises from the same antiunitary mechanism.

We now apply Eq.~\eqref{eq:selection} directly to quantum error correction. A set $\{E_a\}$ is correctable exactly when it satisfies the Knill-Laflamme condition \cite{KnillLaflamme}
\begin{equation}
 PE_a^\dagger E_bP=c_{ab}P
 \label{eq:KLpairs}
\end{equation}
for every pair. For Pauli errors of weight at most $t$, each product in Eq.~\eqref{eq:KLpairs} is, up to phase, a Hermitian Pauli of weight at most $2t$. Kramers symmetry supplies every even-weight condition. Only odd weights $1,3,\ldots,2t-1$ remain to be imposed.

For $t=1$, the conclusion is especially sharp. Products of errors on distinct qubits have weight two and are automatic. Products on the same qubit are the identity or, up to phase, another weight-one Pauli already covered by detection. Hence
\begin{equation}
 \boxed{\Theta\calC=\calC,\quad d\geq2
 \quad\Longrightarrow\quad d\geq3.}
 \label{eq:detectcorrect}
\end{equation}
This is stronger than saying that each one-qubit operator has vanishing off-diagonal matrix elements. Correction requires all cross terms $E_a^\dagger E_b$, including pairs on different qubits. Time reversal supplies precisely those missing weight-two products. Equivalently, every Kramers code has odd distance: if $d\geq2$, then $d\geq3$.

More generally, a generic code with distance $d=2t+1$ must satisfy the Knill-Laflamme conditions for all $\sum_{w=1}^{2t}3^w\binom nw$ many Pauli errors of weight $2t$ or less. In the Kramers class all even weights disappear; for fixed $t$, the leading error count drops from  $O(n^{2t})$ to $O(n^{2t-1})$. At distance three this is the reduction
\begin{equation}
 3n+9\binom n2\quad\longrightarrow\quad3n.
 \label{eq:constraints}
\end{equation}
It is therefore natural to work directly with Kramers pairs
\begin{equation}
 \calC_\psi=\operatorname{span}\{|\psi\rangle,\Theta|\psi\rangle\}.
 \label{eq:kramersansatz}
\end{equation}
Kramers orthogonality, $\langle\psi|\Theta\psi\rangle=0$, makes Eq.~\eqref{eq:kramersansatz} a logical qubit for every normalized $|\psi\rangle$ on odd $n$. If such a pair satisfies the KL conditions for the odd weight errors of weight less than $ d $ then the code automatically has distance $ d $.

Our earlier real, exactly $X/Z$-transversal construction of Ref.~\cite{KubischtaReal} is a sufficient route to Kramers invariance, but not the source of the effect. If $P$ is real and $X^{\otimes n},Z^{\otimes n}$ realize logical $X_L,Z_L$, then $Y^{\otimes n}\overline P Y^{\otimes n}=P$. Conversely, arbitrary local $\SU(2)$ rotations commute with $\Theta$ and generate continuous families of complex, nontransversal Kramers codes with the same distance. The intrinsic hypothesis is invariance of the subspace, not reality or a preferred transversal presentation.

The same conclusion can be read term by term in the Shor--Laflamme enumerators \cite{ShorLaflamme}. For a Hermitian Pauli $E$, define
\begin{align}
    A(E) &= \frac{1}{(\dim \C)^2} |\Tr(PE)|^2, \\ B(E) &=\frac{1}{\dim \C} \Tr(PEPE). 
\end{align}
The weight-$w$ enumerator coefficients are the error sums $A_w=\sum_{\wt(E)=w}A(E)$ and $B_w=\sum_{\wt(E)=w}B(E)$. Equation~\eqref{eq:logicaldecomp} gives, without summing over errors,
\begin{equation}
 A(E)=a_0^2,\qquad B(E)-A(E)=\norm{\bm a}^2.
 \label{eq:ABlogical}
\end{equation}
Thus $A(E)$ is the squared common-mode component of the error and $B(E)-A(E)$ is the squared norm of its logical Pauli component. Equation~\eqref{eq:selection} says, assuming time reversal symmetry, that when $\wt(E)$ is even then $\bm{a} = \bm{0}$ and when $\wt(E)$ is odd then $a_0 = 0$. This forces $A(E)=B(E)$ for each even-weight error and $A(E)=0$ for each odd-weight error. This means the Knill-Laflamme condition is satisfied for all even weight errors, even those errors with weight exceeding the distance of the code. If the code $\C$ is a stabilizer code, then the condition $A(E) = 0$ for odd weight errors means there are no odd-weight stabilizers. 

\emph{Shadows are time-reversed codes.---}
For any Hermitian operator $M$, Rains defined the spin flip \cite{RainsShadow}
\begin{equation}
 \widetilde M=Y^{\otimes n}\overline M Y^{\otimes n}
 \label{eq:rainsflip}
\end{equation}
and used it to define the weight-$w$ quantum shadow enumerator coefficients
\begin{align}
    \mathcal{S}_w = \frac{1}{\dim \C} \sum_{\wt(E) = w} \Tr(P E \widetilde{P} E).
\end{align}
Rains used the shadow to strengthen linear-programming bounds for quantum codes; notably, these bounds rule out a three-qubit error-detecting code (see \cite{GottesmanSurviving}).

Whereas the $A$ and $B$ weight enumerators have natural interpretations (namely, they count the number of stabilizers and normalizers for stabilizer codes), the shadow has no comparably direct interpretation, even for stabilizer codes. We remedy that deficiency here. 

We start by noticing that the spin flip operation $\widetilde{M}$ is just conjugation by time-reversal symmetry, that is,
\begin{align}
    \widetilde{M} = \Theta M \Theta^{-1}. 
\end{align}
Thus $\widetilde{P}$ is the projector onto the time-reversed code $\Theta \C$. It follows that $\mathcal{S}_w$ is the sum of overlaps between the time-reversed code and an error-translate of $\C$.

Rains called a code even when $P=\widetilde P$ and odd when $\Tr(P\widetilde P)=0$ \cite{RainsShadow}. Because $\widetilde{P}$ is the projector onto the time-reversed code, Rains-even means the code is time-reversal invariant and Rains-odd means the code and its time-reversed image are orthogonal: 
\begin{equation}
 \begin{aligned}
 \text{Rains-even}&\iff\Theta\calC=\calC,\\
 \text{Rains-odd}&\iff\calC\perp\Theta\calC.
 \end{aligned}
 \label{eq:evenodd}
\end{equation}
Codes between these extremes carry a quantitative symmetry defect,
\begin{align}
    \mathcal S_0(P) &=\frac1{\dim \C}\Tr(P\widetilde P), \\
 \norm{P-\widetilde P}_2^2 &=2(\dim \C)\qty( 1-\mathcal S_0(P)). 
\end{align}
In particular, $\mathcal{S}_0$ asks whether the code coincides with its time-reverse,  and $1- \mathcal{S}_0$ is proportional to a squared distance between $\C$ and its time-reversed image $\Theta \C$.

\emph{Beyond spin-$1/2$.---}
The Pauli-weight rule is a special case of a more general Kramers error-selection principle \cite{KubischtaThesis}. Consider a single spin-$j$ Hilbert space $\mathcal{H}_j$, with physical time reversal
\begin{equation}
    \Theta_j
    =
    e^{-i\pi J_y}K,
    \qquad
    \Theta_j^2=(-1)^{2j},
    \qquad
    \Theta_j J_\alpha \Theta_j^{-1}=-J_\alpha .
    \label{eq:spinTR}
\end{equation}
Under conjugation by $\mathrm{SU}(2)$, its operator space decomposes into irreducible spherical-tensor subspaces,
\begin{equation}
    \operatorname{End}(\mathcal{H}_j)
    =
    \bigoplus_{k=0}^{2j}\mathcal{T}_k,
\end{equation}
where each subspace $\mathcal{T}_k$ is spanned by the Hermitian symmetrized spherical-tensors $T_q^k$ for $-k \leq q \leq k$ satisfying
\begin{equation}
    \left(T_q^k\right)^\dagger=T_q^k,
    \qquad
    \Theta_j T_q^k \Theta_j^{-1}
    =
    (-1)^k T_q^k .
    \label{eq:sphericalTR}
\end{equation}
Thus tensors of odd rank are time-reversal odd, while tensors of even rank are time-reversal even. If $j$ is half-integer and $P$ projects onto a time-reversal-invariant logical qubit, the code is again a Kramers doublet, and Eq.~\eqref{eq:generalrule} gives
\begin{align}
    k\ \text{even}:
    &\qquad
    P T_q^k P
    =
    \frac{\operatorname{Tr}\left(P T_q^k\right)}{2}P,
     \\
    k\ \text{odd}:
    &\qquad
    \operatorname{Tr}\left(P T_q^k P\right)=0 .
    \label{eq:sphericalSelection}
\end{align}
Spherical-tensor rank parity therefore plays the role of Pauli-weight parity.

This immediately has an error-correction consequence. The rank-one tensors span the linear spin operators,
\begin{equation}
    \operatorname{span}
    \left\{
        T_{-1}^1,T_0^1,T_1^1
    \right\}
    =
    \operatorname{span}
    \left\{
        J_x,J_y,J_z
    \right\},
\end{equation}
and describe infinitesimal rotations or magnetic-field noise. Suppose the code detects these errors,
\begin{equation}
    P T_q^1 P=0,
    \qquad
    q=-1,0,1.
    \label{eq:rankOneDetection}
\end{equation}
The product of two rank-one tensors decomposes into tensors of ranks zero, one, and two:
\begin{equation}
    \left(T_q^1\right)^\dagger T_{q'}^1
    \in
    \mathcal{T}_0
    \oplus
    \mathcal{T}_1
    \oplus
    \mathcal{T}_2 .
    \label{eq:rankOneProduct}
\end{equation}
The rank-zero and rank-two components are time-reversal even and hence act as scalars on the code, while the rank-one component vanishes upon compression by Eq.~\eqref{eq:rankOneDetection}, see \cite{gross1,gross2,us1,us2}. Therefore
\begin{equation}
    P
    \left(T_q^1\right)^\dagger
    T_{q'}^1
    P
    =
    c_{qq'}P,
\end{equation}
so the full Knill--Laflamme conditions hold for
\begin{equation}
    \operatorname{span}
    \left\{
        I,T_{-1}^1,T_0^1,T_1^1
    \right\}
    =
    \operatorname{span}
    \left\{
        I,J_x,J_y,J_z
    \right\}.
\end{equation}
Thus detection of all linear-spin errors is sufficient for their correction.

The multiqubit and single-spin settings are two endpoints of the same construction. A heterogeneous register $\bigotimes_a\mathcal{H}_{j_a}$ is Kramers whenever it contains an odd number of half-integer-spin constituents. Under product time reversal, a product of local spherical tensors has parity $(-1)^{\sum_a k_a}$. Any invariant logical qubit therefore inherits a selection rule graded by total tensor rank: even-total-rank errors act as scalars, while odd-total-rank errors have traceless compression. Spin-$1/2$ arrays recover Pauli-weight parity, a single spin recovers tensor-rank parity, and arbitrary mixtures interpolate between them.

Taken together, these results identify time reversal as a structural principle for quantum error correction, rather than merely a symmetry of a protecting Hamiltonian. For a Kramers logical qubit, the antiunitary grading removes an entire parity sector of the Knill--Laflamme constraints: this becomes Pauli-weight parity for spin-$1/2$ registers and spherical-tensor-rank parity for a single spin, with heterogeneous spin registers interpolating between them. For qubit codes, the same structure also underlies the Rains shadow: the spin flip is precisely time reversal, so the shadow coefficients are error-resolved overlaps with the time-reversed code. Thus the antiunitary structure responsible for Kramers protection simultaneously organizes the error-correction conditions and the shadow invariants of a quantum code.


\emph{Acknowledgments.---}  During the preparation of this manuscript the authors used ChatGPT 5.6-Sol and Claude Opus 4.8 to assist with readability and exposition. The authors reviewed and verified all scientific content and take full responsibility for the manuscript.

\bibliography{biblio}

\begin{thebibliography}{15}%
\makeatletter
\providecommand \@ifxundefined [1]{%
 \@ifx{#1\undefined}
}%
\providecommand \@ifnum [1]{%
 \ifnum #1\expandafter \@firstoftwo
 \else \expandafter \@secondoftwo
 \fi
}%
\providecommand \@ifx [1]{%
 \ifx #1\expandafter \@firstoftwo
 \else \expandafter \@secondoftwo
 \fi
}%
\providecommand \natexlab [1]{#1}%
\providecommand \enquote  [1]{``#1''}%
\providecommand \bibnamefont  [1]{#1}%
\providecommand \bibfnamefont [1]{#1}%
\providecommand \citenamefont [1]{#1}%
\providecommand \href@noop [0]{\@secondoftwo}%
\providecommand \href [0]{\begingroup \@sanitize@url \@href}%
\providecommand \@href[1]{\@@startlink{#1}\@@href}%
\providecommand \@@href[1]{\endgroup#1\@@endlink}%
\providecommand \@sanitize@url [0]{\catcode `\\12\catcode `\$12\catcode `\&12\catcode `\#12\catcode `\^12\catcode `\_12\catcode `\%12\relax}%
\providecommand \@@startlink[1]{}%
\providecommand \@@endlink[0]{}%
\providecommand \url  [0]{\begingroup\@sanitize@url \@url }%
\providecommand \@url [1]{\endgroup\@href {#1}{\urlprefix }}%
\providecommand \urlprefix  [0]{URL }%
\providecommand \Eprint [0]{\href }%
\providecommand \doibase [0]{https://doi.org/}%
\providecommand \selectlanguage [0]{\@gobble}%
\providecommand \bibinfo  [0]{\@secondoftwo}%
\providecommand \bibfield  [0]{\@secondoftwo}%
\providecommand \translation [1]{[#1]}%
\providecommand \BibitemOpen [0]{}%
\providecommand \bibitemStop [0]{}%
\providecommand \bibitemNoStop [0]{.\EOS\space}%
\providecommand \EOS [0]{\spacefactor3000\relax}%
\providecommand \BibitemShut  [1]{\csname bibitem#1\endcsname}%
\let\auto@bib@innerbib\@empty
\bibitem [{\citenamefont {Wigner}(1959)}]{Wigner}%
  \BibitemOpen
  \bibfield  {author} {\bibinfo {author} {\bibfnamefont {E.~P.}\ \bibnamefont {Wigner}},\ }\href@noop {} {\emph {\bibinfo {title} {Group Theory and Its Application to the Quantum Mechanics of Atomic Spectra}}}\ (\bibinfo  {publisher} {Academic Press},\ \bibinfo {address} {New York},\ \bibinfo {year} {1959})\BibitemShut {NoStop}%
\bibitem [{\citenamefont {Kramers}(1930)}]{Kramers}%
  \BibitemOpen
  \bibfield  {author} {\bibinfo {author} {\bibfnamefont {H.~A.}\ \bibnamefont {Kramers}},\ }\bibfield  {title} {\bibinfo {title} {Th\'eorie g\'en\'erale de la rotation paramagn\'etique dans les cristaux},\ }\href@noop {} {\bibfield  {journal} {\bibinfo  {journal} {Proc. Amsterdam Acad.}\ }\textbf {\bibinfo {volume} {33}},\ \bibinfo {pages} {959} (\bibinfo {year} {1930})}\BibitemShut {NoStop}%
\bibitem [{\citenamefont {Dyson}(1962)}]{Dyson}%
  \BibitemOpen
  \bibfield  {author} {\bibinfo {author} {\bibfnamefont {F.~J.}\ \bibnamefont {Dyson}},\ }\bibfield  {title} {\bibinfo {title} {The threefold way: Algebraic structure of symmetry groups and ensembles in quantum mechanics},\ }\href@noop {} {\bibfield  {journal} {\bibinfo  {journal} {J. Math. Phys.}\ }\textbf {\bibinfo {volume} {3}},\ \bibinfo {pages} {1199} (\bibinfo {year} {1962})}\BibitemShut {NoStop}%
\bibitem [{\citenamefont {Kane}\ and\ \citenamefont {Mele}(2005)}]{KaneMele}%
  \BibitemOpen
  \bibfield  {author} {\bibinfo {author} {\bibfnamefont {C.~L.}\ \bibnamefont {Kane}}\ and\ \bibinfo {author} {\bibfnamefont {E.~J.}\ \bibnamefont {Mele}},\ }\bibfield  {title} {\bibinfo {title} {{$Z_2$} topological order and the quantum spin hall effect},\ }\href@noop {} {\bibfield  {journal} {\bibinfo  {journal} {Phys. Rev. Lett.}\ }\textbf {\bibinfo {volume} {95}},\ \bibinfo {pages} {146802} (\bibinfo {year} {2005})}\BibitemShut {NoStop}%
\bibitem [{\citenamefont {Fu}\ \emph {et~al.}(2007)\citenamefont {Fu}, \citenamefont {Kane},\ and\ \citenamefont {Mele}}]{FuKaneMele}%
  \BibitemOpen
  \bibfield  {author} {\bibinfo {author} {\bibfnamefont {L.}~\bibnamefont {Fu}}, \bibinfo {author} {\bibfnamefont {C.~L.}\ \bibnamefont {Kane}},\ and\ \bibinfo {author} {\bibfnamefont {E.~J.}\ \bibnamefont {Mele}},\ }\bibfield  {title} {\bibinfo {title} {Topological insulators in three dimensions},\ }\href@noop {} {\bibfield  {journal} {\bibinfo  {journal} {Phys. Rev. Lett.}\ }\textbf {\bibinfo {volume} {98}},\ \bibinfo {pages} {106803} (\bibinfo {year} {2007})}\BibitemShut {NoStop}%
\bibitem [{\citenamefont {Knill}\ and\ \citenamefont {Laflamme}(1997)}]{KnillLaflamme}%
  \BibitemOpen
  \bibfield  {author} {\bibinfo {author} {\bibfnamefont {E.}~\bibnamefont {Knill}}\ and\ \bibinfo {author} {\bibfnamefont {R.}~\bibnamefont {Laflamme}},\ }\bibfield  {title} {\bibinfo {title} {Theory of quantum error-correcting codes},\ }\href@noop {} {\bibfield  {journal} {\bibinfo  {journal} {Phys. Rev. A}\ }\textbf {\bibinfo {volume} {55}},\ \bibinfo {pages} {900} (\bibinfo {year} {1997})}\BibitemShut {NoStop}%
\bibitem [{\citenamefont {Rains}(1999)}]{RainsShadow}%
  \BibitemOpen
  \bibfield  {author} {\bibinfo {author} {\bibfnamefont {E.~M.}\ \bibnamefont {Rains}},\ }\bibfield  {title} {\bibinfo {title} {Quantum shadow enumerators},\ }\href@noop {} {\bibfield  {journal} {\bibinfo  {journal} {IEEE Trans. Inf. Theory}\ }\textbf {\bibinfo {volume} {45}},\ \bibinfo {pages} {2361} (\bibinfo {year} {1999})}\BibitemShut {NoStop}%
\bibitem [{\citenamefont {Kubischta}\ \emph {et~al.}(2023)\citenamefont {Kubischta}, \citenamefont {Teixeira},\ and\ \citenamefont {Silvester}}]{KubischtaReal}%
  \BibitemOpen
  \bibfield  {author} {\bibinfo {author} {\bibfnamefont {E.}~\bibnamefont {Kubischta}}, \bibinfo {author} {\bibfnamefont {I.}~\bibnamefont {Teixeira}},\ and\ \bibinfo {author} {\bibfnamefont {J.~M.}\ \bibnamefont {Silvester}},\ }\href@noop {} {\bibinfo {title} {Quantum weight enumerators for real codes with {$X$} and {$Z$} exactly transversal}} (\bibinfo {year} {2023}),\ \Eprint {https://arxiv.org/abs/2306.12526} {arXiv:2306.12526 [quant-ph]} \BibitemShut {NoStop}%
\bibitem [{\citenamefont {Shor}\ and\ \citenamefont {Laflamme}(1997)}]{ShorLaflamme}%
  \BibitemOpen
  \bibfield  {author} {\bibinfo {author} {\bibfnamefont {P.~W.}\ \bibnamefont {Shor}}\ and\ \bibinfo {author} {\bibfnamefont {R.}~\bibnamefont {Laflamme}},\ }\bibfield  {title} {\bibinfo {title} {Quantum analog of the macwilliams identities for classical coding theory},\ }\href@noop {} {\bibfield  {journal} {\bibinfo  {journal} {Phys. Rev. Lett.}\ }\textbf {\bibinfo {volume} {78}},\ \bibinfo {pages} {1600} (\bibinfo {year} {1997})}\BibitemShut {NoStop}%
\bibitem [{\citenamefont {Gottesman}(2026)}]{GottesmanSurviving}%
  \BibitemOpen
  \bibfield  {author} {\bibinfo {author} {\bibfnamefont {D.}~\bibnamefont {Gottesman}},\ }\href@noop {} {\bibinfo {title} {Surviving as a quantum computer in a classical world}},\ \bibinfo {howpublished} {\url{https://www.cs.umd.edu/~dgottesm/}} (\bibinfo {year} {2026}),\ \bibinfo {note} {draft textbook, 2026 draft}\BibitemShut {NoStop}%
\bibitem [{\citenamefont {Kubischta}(2025)}]{KubischtaThesis}%
  \BibitemOpen
  \bibfield  {author} {\bibinfo {author} {\bibfnamefont {E.}~\bibnamefont {Kubischta}},\ }\emph {\bibinfo {title} {Quantum Codes from Symmetry}},\ \href@noop {} {Ph.D. thesis},\ \bibinfo  {school} {University of Maryland, College Park} (\bibinfo {year} {2025})\BibitemShut {NoStop}%
\bibitem [{\citenamefont {Gross}(2021)}]{gross1}%
  \BibitemOpen
  \bibfield  {author} {\bibinfo {author} {\bibfnamefont {J.~A.}\ \bibnamefont {Gross}},\ }\bibfield  {title} {\bibinfo {title} {Designing codes around interactions: The case of a spin},\ }\href {https://doi.org/10.1103/PhysRevLett.127.010504} {\bibfield  {journal} {\bibinfo  {journal} {Phys. Rev. Lett.}\ }\textbf {\bibinfo {volume} {127}},\ \bibinfo {pages} {010504} (\bibinfo {year} {2021})}\BibitemShut {NoStop}%
\bibitem [{\citenamefont {Omanakuttan}\ and\ \citenamefont {Gross}(2023)}]{gross2}%
  \BibitemOpen
  \bibfield  {author} {\bibinfo {author} {\bibfnamefont {S.}~\bibnamefont {Omanakuttan}}\ and\ \bibinfo {author} {\bibfnamefont {J.~A.}\ \bibnamefont {Gross}},\ }\bibfield  {title} {\bibinfo {title} {Multispin clifford codes for angular momentum errors in spin systems},\ }\href {https://doi.org/10.1103/PhysRevA.108.022424} {\bibfield  {journal} {\bibinfo  {journal} {Phys. Rev. A}\ }\textbf {\bibinfo {volume} {108}},\ \bibinfo {pages} {022424} (\bibinfo {year} {2023})}\BibitemShut {NoStop}%
\bibitem [{\citenamefont {Kubischta}\ and\ \citenamefont {Teixeira}(2023)}]{us1}%
  \BibitemOpen
  \bibfield  {author} {\bibinfo {author} {\bibfnamefont {E.}~\bibnamefont {Kubischta}}\ and\ \bibinfo {author} {\bibfnamefont {I.}~\bibnamefont {Teixeira}},\ }\bibfield  {title} {\bibinfo {title} {Family of quantum codes with exotic transversal gates},\ }\href {https://doi.org/10.1103/PhysRevLett.131.240601} {\bibfield  {journal} {\bibinfo  {journal} {Phys. Rev. Lett.}\ }\textbf {\bibinfo {volume} {131}},\ \bibinfo {pages} {240601} (\bibinfo {year} {2023})}\BibitemShut {NoStop}%
\bibitem [{\citenamefont {Kubischta}\ and\ \citenamefont {Teixeira}(2025)}]{us2}%
  \BibitemOpen
  \bibfield  {author} {\bibinfo {author} {\bibfnamefont {E.}~\bibnamefont {Kubischta}}\ and\ \bibinfo {author} {\bibfnamefont {I.}~\bibnamefont {Teixeira}},\ }\bibfield  {title} {\bibinfo {title} {Permutation-invariant quantum codes with transversal generalized phase gates},\ }\href {https://doi.org/10.1109/TIT.2024.3487964} {\bibfield  {journal} {\bibinfo  {journal} {IEEE Transactions on Information Theory}\ }\textbf {\bibinfo {volume} {71}},\ \bibinfo {pages} {485} (\bibinfo {year} {2025})}\BibitemShut {NoStop}%
\end{thebibliography}%

\end{document}